\documentclass[a4paper,12pt]{article}
\pdfoutput=1
\PassOptionsToPackage{hypertexnames=false}{hyperref}

\usepackage{jheppub}
\usepackage[T1]{fontenc}
\usepackage{bm}
\usepackage{cancel}
\usepackage{makecell}
\usepackage{placeins}
\numberwithin{equation}{section}

\title{Bayesian analysis of the complex singlet model with phase transition gravitational waves}

\author[a,b]{Qingyuan Liang,}
\author[c]{Ligong Bian,}
\author[a,b]{Huai-Ke Guo,}
\author[d,e]{and Yongcheng Wu}

\affiliation[a]{International Centre for Theoretical Physics Asia-Pacific (ICTP-AP), University of Chinese Academy of Sciences (UCAS), Beijing, China}
\affiliation[b]{Taiji Laboratory for Gravitational Wave Universe (Beijing/Hangzhou), University of Chinese Academy of Sciences (UCAS), Beijing, China}
\affiliation[c]{Department of Physics and Chongqing Key Laboratory for Strongly Coupled Physics, Chongqing University, Chongqing 401331, People's Republic of China}
\affiliation[d]{Department of Physics, Institute of Theoretical Physics and Institute of Physics Frontiers and Interdisciplinary Sciences, Nanjing Normal University, Nanjing 210023, China}
\affiliation[e]{Nanjing Key Laboratory of Particle Physics and Astrophysics, Nanjing 210023, China}

\emailAdd{lgbycl@cqu.edu.cn}
\emailAdd{guohuaike@ucas.ac.cn}
\emailAdd{ycwu@njnu.edu.cn}

\abstract{We explore the prospects of probing the complex singlet extension of the Standard Model (CxSM) with gravitational waves from the electroweak phase transition. The study establishes a connection of the scalar potential parameters, the thermodynamic properties of the phase transition, with the directly measured stochastic gravitational-wave background in the presence of astrophysical background and foreground. Considering the space-based gravitational-wave detector Taiji, we construct a frequency-domain likelihood that incorporates instrumental and astrophysical noises, and we perform both Fisher-matrix forecasts and Bayesian nested sampling analysis. The comparison of these two approaches demonstrates consistent parameter recovery and highlights the sensitivity of Taiji to millihertz gravitational-wave signals. We further propagate the inferred constraints on the gravitational-wave spectrum back to the underlying CxSM parameters, obtaining meaningful limits on the Higgs self-couplings. The results emphasize the complementarity between gravitational-wave observations and collider measurements, showing that future missions such as Taiji can serve as a powerful probe of electroweak-scale new physics and the dynamical origin of the Higgs sector.}

\begin{document}

\maketitle

\section{Introduction}

The stochastic gravitational-wave background (SGWB) encodes relic information from the earliest stages of the Universe and provides a unique avenue for probing fundamental physics beyond the reach of colliders~\cite{Caldwell:2022qsj,LiGong:2024qmt}. Evidence for such a background has recently been reported in the nanohertz band by pulsar timing array collaborations~\cite{NANOGrav:2023gor,NANOGrav:2023hde,EPTA:2023sfo, Athron:2023mer}. At higher frequencies, ground-based detectors such as Advanced LIGO~\cite{LIGOScientific:2017vtl}, Virgo~\cite{Kumar:2024bfe}, and KAGRA~\cite{KAGRA:2020tym} have constrained the SGWB up to the kilohertz regime~\cite{LIGOScientific:2025bgj}. In the coming decade, space-based interferometers, including Laser Interferometer Space Antenna (LISA)~\cite{LISA:2017pwj,Robson:2018ifk,LISACosmologyWorkingGroup:2022jok}, Taiji~\cite{Hu:2017mde,Ruan:2018tsw,Wu:2018clg}, and TianQin~\cite{TianQin:2015yph,TianQin:2020hid,Luo:2020bls}, are expected to open an observational window in the millihertz band, enabling tests of a wide range of cosmological processes.

Among the potential cosmological sources, first-order phase transitions (FOPTs) are of particular interest~\cite{Caprini:2015zlo,Weir:2017wfa,Mazumdar:2018dfl,Caprini:2019egz,Bian:2021ini,Athron:2023xlk}. The violent bubble expansion and collision during such transitions can generate a stochastic gravitational-wave (GW) signal through sound waves (SWs) and plasma turbulence. A strongly first-order electroweak phase transition (EWPT) can simultaneously satisfy the out-of-equilibrium condition required for electroweak baryogenesis~\cite{Morrissey:2012db} and yield a potentially detectable GW signal in the frequency range accessible to future space missions~\cite{Caprini:2015zlo}.

The extraction of a faint cosmological signal from experimental data is, however, complicated by instrumental noise and astrophysical confusion foregrounds~\cite{Boileau:2020rpg,Boileau:2021sni}. Space-based interferometers rely on time-delay interferometry (TDI)~\cite{Tinto:2014lxa} to suppress laser frequency fluctuations, forming three orthogonal data combinations, commonly labeled as $A$, $E$, and $T$~\cite{Smith:2019wny}. The so-called null channel $T$~\cite{Adams:2010vc} is nearly insensitive to GWs at low frequencies, serving as an internal reference for calibration and noise validation. Foregrounds originating from unresolved compact binaries, both Galactic and extragalactic, constitute an additional stochastic background~\cite{Tinto:2004wu,Barack:2004wc,Caprini:2024hue}, which must be carefully modeled and marginalized in any realistic analysis.

Recent developments have emphasized the importance of combining theoretical modeling with statistically rigorous data analysis~\cite{Gowling:2021gcy,Gowling:2022pzb,Boileau:2022ter,Caprini:2024hue,Lewicki:2024xan,Huang:2025uer}. Following this direction, we construct a frequency-domain framework for parameter inference that incorporates realistic noise spectra, astrophysical foregrounds, and the detector response of a Taiji-like mission. Both the Fisher information matrix (FIM) forecasts and full Bayesian analysis are performed to evaluate the reconstruction accuracy of the SGWB parameters and to map these spectral constraints onto the underlying particle-physics parameters. Compared with our previous study~\cite{Guan:2025idx}, the present work further accounts for the astrophysical foreground, which significantly affects the detectability of the cosmological background and provides a more realistic assessment. In addition, we employ the nested sampling (NS) algorithm implemented in \texttt{bilby.dynesty}~\cite{Ashton:2018jfp} instead of \texttt{PyMC}, enabling GPU acceleration~\cite{Prathaban:2025qgg} and providing direct estimates of the Bayes factor (BF) for model comparison.

The interplay between collider measurements and GW observations has emerged as a promising frontier for probing the structure of the Higgs potential~\cite{Ramsey-Musolf:2019lsf}. In particular, the cubic and quartic Higgs self-couplings, which remain challenging to measure directly at colliders~\cite{DiVita:2017vrr,Zabinski:2023jhr}, can be indirectly constrained through GWs produced during the EWPT~\cite{Alves:2018jsw}. Building upon our previous work~\cite{Guo:2023koq}, we further investigate the complex singlet extension of the Standard Model (CxSM)~\cite{Barger:2008jx, Ghosh:2025rbt}, aiming to provide deeper insights into the connection between GW observations and particle-physics phenomenology~\cite{Biekotter:2025npc}. The model capitalizes on the multistep phase transition that provides the chance to address dark matter and baryogenesis simultaneously~\cite{Jiang:2015cwa,Chiang:2017nmu}.

Using simulated data for a Taiji-like interferometer, we evaluate the detectability of the phase-transition-induced GW spectrum, quantify the expected parameter precision, and propagate the inferred uncertainties to the CxSM scalar potential parameters. The resulting bounds \cite{Chen:2019ebq} on the Higgs self-couplings demonstrate the complementarity between GW and collider probes of the electroweak symmetry-breaking mechanism.

The remainder of this paper is organized as follows. Section~\ref{jvk} introduces the CxSM framework and its scalar potential. Section~\ref{vdf} presents the Taiji detector response and the statistical analysis pipeline. Section~\ref{vsm} details the parameter estimation results and their implications for the underlying particle-physics model. Conclusions are presented in Sec.~\ref{qmg}.

\section{Complex Singlet Extension of the Standard Model}\label{jvk}

In this work, we consider the SM extended by a complex scalar singlet $\mathbb{S}$ as a benchmark scenario. The most general scalar potential for such an extension is given as \cite{Chen:2019ebq, Barger:2008jx}
\begin{align}
    V(\Phi,\mathbb{S}) ={}& \mu^2|\Phi|^2 + \lambda |\Phi|^4
    + \frac{\delta_2}{2}|\Phi|^2|\mathbb{S}|^2
    + \frac{b_2}{2}|\mathbb{S}|^2 + \frac{d_2}{4}|\mathbb{S}|^4 \nonumber\\
    &+ \left(\frac{\delta_1}{4}|\Phi|^2\mathbb{S}
    + \frac{\delta_3}{4}|\Phi|^2\mathbb{S}^2 + c.c.\right) \nonumber\\
    &+ \left(a_1 \mathbb{S} + \frac{b_1}{4}\mathbb{S}^2
    + \frac{c_1}{6}\mathbb{S}^3 + \frac{c_2}{6}\mathbb{S}|\mathbb{S}|^2 \right. \nonumber\\
    &\left.\hspace{1.4cm}+ \frac{d_1}{8}\mathbb{S}^4
    + \frac{d_3}{8}\mathbb{S}^2|\mathbb{S}|^2 + c.c.\right), \nonumber
\end{align}
where $\Phi$ is the SM scalar doublet. The parameters in the first line of the potential are real, while the parameters in the second line is generally complex. There are two possible global symmetries that can be imposed on the potential: (i) a discrete $\mathbb{Z}_2$ symmetry under which $\mathbb{S}\to -\mathbb{S}$ can be imposed to eliminate all terms with odd powers of $\mathbb{S}$, e.g., $\delta_1, a_1, c_{1,2}$; (ii) a global $U(1)$ symmetry under which $\mathbb{S}\to e^{i\alpha}\mathbb{S}$ eliminates all terms with complex coefficients, e.g., $\delta_{1,3},a_1,b_1,c_{1,2},d_{1,3}$. In the current study, we consider the case where $\mathbb{S}$ acquires a nonzero vacuum expectation value (VEV) at zero temperature. As a consequence, the real component of $\mathbb{S}$ mixes with the SM Higgs. $U(1)$ and $\mathbb{Z}_2$ symmetries are both spontaneously broken by the singlet VEV, and the Goldstone boson corresponding to the single $\mathbb{S}$ is stable but massless. A $U(1)$ symmetry-breaking term is required to generate a corresponding mass, for which the $b_1$ term is used. On the other hand, the spontaneously broken $\mathbb{Z}_2$ symmetry will lead to the cosmological domain wall problem \cite{Zeldovich:1974uw, Friedland:2002qs}. To avoid this, we further introduce a soft $\mathbb{Z}_2$-breaking term ($a_1$). Then, the potential we considered for the following study reads \cite{Barger:2008jx}
\begin{align} 
    V(\Phi,\mathbb{S}) =& \mu^2|\Phi|^2 + \lambda |\Phi|^4 + \frac{\delta_2}{2}|\Phi|^2|\mathbb{S}|^2 + \frac{b_2}{2}|\mathbb{S}|^2 + \frac{d_2}{4}|\mathbb{S}|^4\nonumber\\
    &\quad + \left(a_1 \mathbb{S} + \frac{b_1}{4}\mathbb{S}^2 + c.c.\right).
\end{align}
Note that this is not a unique choice for the potential purpose discussed above.

To minimize the scalar potential, we expand the scalar around the VEVs as
\begin{align}
    \Phi = \begin{pmatrix}
        0\\
        \frac{v+\phi}{\sqrt{2}}
    \end{pmatrix},\quad \mathbb{S} = \frac{v_s + \eta + iA}{\sqrt{2}},
\end{align}
where we have suppressed the Goldstone boson. The VEVs are determined by the corresponding minimization conditions as
\begin{equation}
    \begin{cases}
    0 = \frac{\partial V}{\partial v} = \mu^2v+\lambda v^3+\frac{\delta_2}{4}vv_s^2,\\
    0 = \frac{\partial V}{\partial v_s} = \frac{\delta_2}{4}v^2v_s +\sqrt{2}a_1+\frac{b_1+b_2}{2}v_s+\frac{d_2}{4}v_s^3
    \end{cases}
    ,
\end{equation}
thus
\begin{equation}
    \begin{cases}
        \mu^2 = -\lambda v^2 - \frac{\delta_2}{4}v_s^2, \\
        b_1 + b_2 = -2\sqrt{2}\frac{a_1}{v_s} - \frac{\delta_2}{2}v^2 - \frac{d_2}{2}v_s^2
    \end{cases}
    .
\end{equation}
The mass spectrum of the scalars can be obtained after inserting the above relationships as follows:
\begin{align}
    m_A^2 &= -b_1 -\sqrt{2}\frac{a_1}{v_s},\\
    \mathcal{M}_{\phi\eta}^2 &= \begin{pmatrix}
        \mu_\phi^2 & \mu_{\phi\eta}^2 \\
        \mu_{\phi\eta}^2 & \mu_\eta^2
    \end{pmatrix},\\
    \mu_\phi^2 &= 2\lambda v^2, \\
    \mu_\eta^2 &= \frac{d_2}{2}v_s^2 - \sqrt{2}\frac{a_1}{v_s},\\
    \mu_{\phi\eta}^2 &= \frac{\delta_2}{2}vv_s.
\end{align}
The scalars $\phi$ and $\eta$ therefore mix according to
\begin{align}
    \begin{pmatrix} 
        h\\
        s
    \end{pmatrix} &= \begin{pmatrix}
        c_\theta & s_\theta \\
        -s_\theta & c_\theta
    \end{pmatrix}\begin{pmatrix}
        \phi \\
        \eta
    \end{pmatrix},
\end{align}
which diagonalizes the mass matrix with the corresponding masses $m_h$, $m_s$, and $c_\theta \equiv \cos\theta, s_\theta\equiv\sin\theta$. In terms of the masses ($m_h,m_s,m_A$), VEVs ($v,v_s$), and the mixing angle $\theta$, quartic couplings can be expressed as
\begin{align}
    \lambda &= \frac{1}{2v^2}\left(c_\theta^2 m_h^2 + s_\theta^2 m_s^2\right),\\
    d_2 &= \frac{2}{v_s^2}\left(s_\theta^2 m_h^2 + c_\theta^2 m_s^2 + \sqrt{2}\frac{a_1}{v_s}\right),\\
    \delta_2 &= \frac{2}{vv_s}\left(m_h^2-m_s^2\right)s_\theta c_\theta.
\end{align}
Then the input parameters are given as
\begin{align}
    v(=246\,{\rm GeV}),\, v_s,\, m_h(=125\,{\rm GeV}), m_s, m_A, \theta, a_1.
\end{align}
In the following studies, we scan these parameters within the specified ranges,
\begin{equation}
    \begin{aligned}
        &v_s\in[0,150]\,{\rm GeV},\, m_s\in[65,150]\,{\rm GeV},\, m_A\in[65,2000]\,{\rm GeV},\\
        &\theta\in[0,0.5],\, \sqrt[3]{a_1} \in [-100,100]\,{\rm GeV}
    \end{aligned}
\end{equation}
During the scan, every point is also checked against the constraints from the LHC measurements. Those that cannot satisfy the constraints from LHC measurement are ignored. It is worth noting that the electroweak phase transition may occur through either a one-step or a two-step route. However, only a negligible subset of the scanned parameter space exhibits the two-step pattern in our scan, mainly due to the linear term ($a_1$ term) presented in the potential. For this reason, our analysis focuses exclusively on the one-step transition points.

With the above setup for the model parameters, we define the deviations of the cubic and quartic self-couplings of the SM-like Higgs as \cite{Chen:2019ebq}
\begin{align}
    \delta\kappa_3&\equiv \frac{\lambda_{hhh}}{\lambda_{hhh}^{SM}}-1=c_\theta^3 - 1+ \frac{v}{v_s}s_\theta^3+\sqrt{2}\frac{v}{v_s} \frac{a_1}{m_h^2v_s}s_\theta^3 ,\\
    \delta\kappa_4&\equiv \frac{\lambda_{hhhh}}{\lambda_{hhhh}^{SM}}-1 = c_\theta^6-1 + \frac{v^2}{v_s^2}s_\theta^6 + \frac{v^2}{v_s^2}\frac{m_s^2}{m_h^2}s_\theta^4c_\theta^2 + \\
    & 2 \frac{v}{v_s} \frac{m_h^2-m_s^2}{m_h^2}s_\theta^3c_\theta^3 \nonumber
    +\frac{m_s^2}{m_h^2}s_\theta^2c_\theta^4 + \sqrt{2}\frac{v^2}{v_s^2}\frac{a_1}{m_h^2v_s}s_\theta^4 .
\end{align}

\section{The Taiji Detector and Data Analysis Framework} \label{vdf}

Taiji is a proposed space-based GW observatory designed to probe the millihertz frequency band, where a rich variety of astrophysical and cosmological sources are expected to exist. Its triangular constellation, consisting of three drag-free spacecraft in heliocentric orbits, provides multiple long-baseline interferometric measurements that enable high-sensitivity detection of signals such as the SGWB generated by an EWPT. Notably, the predicted peak frequency of the EWPT-induced SGWB typically falls within the optimal sensitivity range of space-based detectors, such as LISA and Taiji.

This section presents an overview of the Taiji mission concept and its interferometric measurement strategy. We describe the TDI technique~\cite{Tinto:2004wu}, which suppresses the otherwise overwhelming laser frequency noise and yields the orthogonal $A$, $E$, and $T$ data channels used for scientific analysis. We then summarize the frequency-domain statistical framework~\cite{Caprini:2024hue} employed in SGWB searches, including the modeling of the detector response functions, instrumental noise spectra, and channel correlations, following the conventions established in~\cite{Guan:2025idx}. Finally, we introduce the likelihood function and the FIM formalism that we use to forecast the constraints on the parameters characterizing the GW spectrum and the underlying particle-physics model.

\subsection{Gravitational waves in space-based detectors} \label{vcn}

Taiji employs a triangular constellation of three spacecraft, labeled $A$, $B$, and $C$, separated by arm lengths of $L = 3\times10^{9}\,\mathrm{m}$ (and $L = 2.5\times10^{9}\,\mathrm{m}$ for LISA)~\cite{Hu:2017mde,Ruan:2018tsw,Wu:2018clg}. Each spacecraft exchanges laser beams with the other two, forming three unequal-arm Michelson interferometers. The resulting frequency-domain data streams $\tilde{d}_i(f)$ ($i=1,2,3$) are linearly recombined into the orthogonal TDI channels $\tilde{d}_I(f)$ ($I = A, E, T$), following the same setup as in~\cite{Guan:2025idx}.

The single-sided power spectral density of each TDI channel is written as
\begin{equation}
    P_a(f) = S_a(f) + N_a(f), \qquad a \in \{A,E,T\},
\end{equation}
where $S_a(f)$ and $N_a(f)$ denote the GW signal and instrumental noise contributions.  
The signal component is related to the GW energy density spectrum via
\begin{equation}\label{tfv}
    S_a(f) =
    \frac{3H_0^2}{4\pi^2}\,
    \frac{\Omega_{\mathrm{GW}}(f)}{f^3}\,
    \mathcal{R}_a(f),
\end{equation}
where $\mathcal{R}_a(f)$ is the detector response. The explicit forms of $\mathcal{R}_a(f)$ and $N_a(f)$ are given in~\cite{Guan:2025idx}.  
We adopt the nominal noise parameters
\[
    N_{\mathrm{acc}} = 3\times10^{-15}, \qquad
    \delta x = 8\times10^{-12},
\]
and $\delta x = 15\times10^{-12}$ for LISA.

Following~\cite{Babak:2021mhe}, the sensitivity function of each TDI channel is defined as
\begin{equation}\label{hgv}
    \mathcal{S}_I(f) = 
    \sqrt{\frac{N_I(f)}{\mathcal{R}_I(f)}},
    \qquad I = A, E, T.
\end{equation}
The corresponding SNR is
\begin{equation}
    \mathrm{SNR}_I =
    \sqrt{
        2T_t
        \int_{0}^{\infty}
        \left( 
            \frac{S_I(f)}{N_I(f)}
        \right)^{\!2}
        \mathrm{d}f
    },
    \qquad I = A, E, T,
\end{equation}
where $T_t$ is the total observation time specified in the next section.

\begin{figure}[t]
    \centering
    \includegraphics[width=0.75\linewidth]{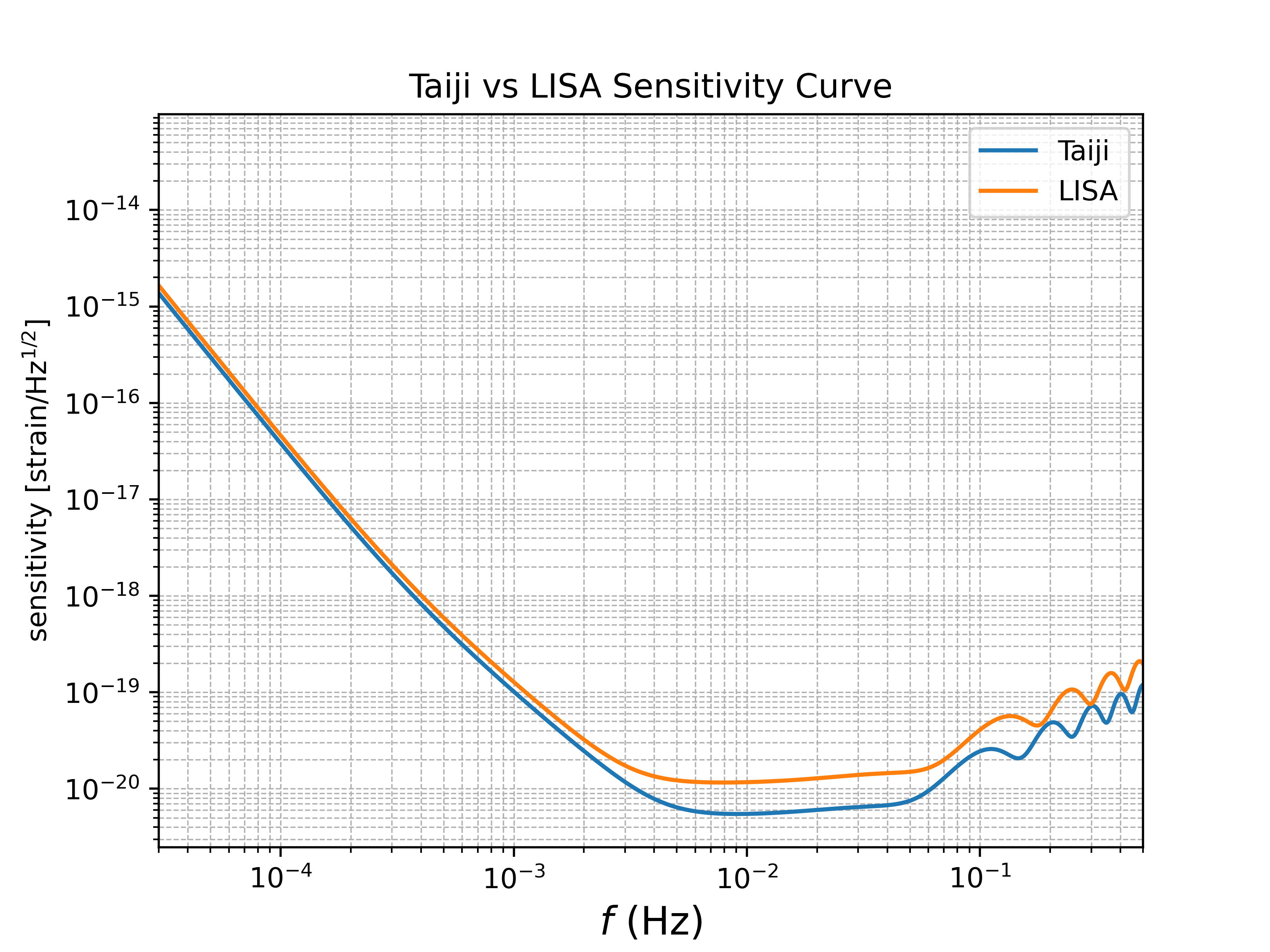}
    \caption{Sensitivity curves for LISA and Taiji in the A or E channel. Solid lines correspond to sensitivities computed from Eq.~\eqref{hgv}.}
    \label{lmn}
\end{figure}

Since the null-channel method assumes that the $T$ channel is free of signals, only the $A/E$ channel is employed for evaluating sensitivity and SNR. The resulting sensitivity curves for LISA and Taiji are shown in Fig.~\ref{lmn}. The solid curves correspond to the model-based sensitivities obtained from Eq.~\eqref{hgv}.

In this work, we consider a mixed scenario in which a cosmological stochastic GW signal is superimposed on both Galactic and extragalactic astrophysical contributions. Following a convention commonly adopted in space-based SGWB analyses (see, e.g., Ref.~\cite{Boileau:2021kiw}), we distinguish these components according to their astrophysical origin and observational properties. In particular, unresolved sources within the Milky Way are referred to as an astrophysical foreground, while unresolved sources of extragalactic origin are referred to as an astrophysical background.

To mitigate the strong parameter degeneracies identified in Refs.~\cite{Caprini:2024hue,Guan:2025idx}, we restrict our analysis of the sound-wave contribution to two geometric parameters that characterize its spectral shape, namely, the peak amplitude $\Omega_0$ and the peak frequency $f_{\mathrm{p}}$.

Under these assumptions, the total GW energy density spectrum is modeled as the sum of three components,
\begin{equation}
    \Omega_{\mathrm{GW}}(f)
    = \Omega_{\mathrm{DWD}}(f)
    + \Omega_{\mathrm{GW,ast}}(f)
    + \Omega_{\mathrm{SW}}(f),
\end{equation}
corresponding to the Galactic astrophysical foreground, the extragalactic astrophysical background, and the cosmological SW signal, respectively.

The Galactic foreground is dominated by the superposition of unresolved double white-dwarf (DWD) binaries within the Milky Way. Because of the spatial structure of the Galaxy and the orbital motion of space-based detectors, this component is anisotropic and exhibits characteristic spectral features. We model it using a broken power-law parametrization~\cite{Chen:2023zkb},
\begin{equation}
    \Omega_{\mathrm{DWD}}(f)
    = \frac{
        A_1 \left( f / f_* \right)^{\alpha_1}
    }{
        1 + A_2 \left( f / f_* \right)^{\alpha_2}
    },
\end{equation}
where $f_*$ denotes a characteristic frequency defined as $f_* = c / (2\pi L)$.

The astrophysical background represents an approximately isotropic stochastic signal generated by the superposition of a large population of unresolved extragalactic compact binaries, such as binary black holes and binary neutron stars at cosmological distances. Rather than modeling individual source populations in detail, we adopt a phenomenological power-law parametrization commonly used in SGWB analyses~\cite{Boileau:2020rpg},
\begin{equation}
    \Omega_{\mathrm{GW,ast}}(f)
    = \Omega_{\mathrm{ast}}
    \left( \frac{f}{f_{\mathrm{ref}}} \right)^{\varepsilon},
\end{equation}
where $\Omega_{\mathrm{ast}}$ is the amplitude at the reference frequency $f_{\mathrm{ref}} = 1~\mathrm{mHz}$ and $\varepsilon$ is the spectral index. In this work, the astrophysical background is treated as a nuisance component, included to assess the robustness of cosmological signal recovery in the presence of realistic astrophysical contamination.

The SW component, corresponding to the acoustic contribution from a FOPT, is modeled by a broken power-law template~\cite{Guan:2025idx,Alves:2018jsw},
\begin{equation}
    \Omega_{\mathrm{SW}}(f)
    = \Omega_0
    \left( \frac{f}{f_{\mathrm{p}}} \right)^3
    \left[
        \frac{7}{4 + 3 \left( f / f_{\mathrm{p}} \right)^2}
    \right]^{7/2}.
\end{equation}

In this study, we focus on assessing the detectability of the cosmological background. Accordingly, the SNR and BF quantify both the strength of the signal and its likelihood of being distinguished from the underlying noise. To characterize the SNR, we consider two definitions. The absolute SNR ($\mathrm{SNR}_a$) compares the cosmological background solely against the instrumental noise, providing a measure of its intrinsic amplitude. In contrast, the relative SNR ($\mathrm{SNR}_r$) incorporates the astrophysical foreground and background into the effective noise budget. This quantity is more relevant for realistic data analysis, as it reflects the observational challenges posed by astrophysical contamination. Moreover, $\mathrm{SNR}_r$ is directly aligned with the interpretation of the BF: both compare a ``signal + noise'' model against a ``noise-only'' model that already includes all astrophysical components, enabling a meaningful and consistent assessment of detectability across different metrics.

\subsection{Likelihood function and Fisher information matrix}

During the operation of space-based detectors, interruptions and data gaps divide the observation period into $N_0$ valid time segments of duration $T$, yielding a total observing time of $T_t = N_0 T$. In this work, we take $T = 10^6~\mathrm{s}$ and $N_0 = 126$, corresponding to an effective duration of approximately four years. The data are sampled at a rate $f_s = 1/\Delta t$, which satisfies $f_s > 2 f_{\max}$ according to the Nyquist criterion, where $f_{\max}$ denotes the highest frequency of interest. The corresponding frequency resolution is $\Delta f = 1/T$. For convenience, we set $\Delta t = 1~\mathrm{s}$, leading to a frequency range of $[3\times10^{-5},\, 0.5]~\mathrm{Hz}$. These conventions are consistent with our previous analysis~\cite{Guan:2025idx} and with the LISA study~\cite{Caprini:2024hue}.

\begin{table}[t!]
    \centering
    \renewcommand{\arraystretch}{1.1}
    \setlength{\tabcolsep}{8pt}
    \resizebox{\textwidth}{!}{%
    \begin{tabular}{|c|c|c|c|c|c|}
        \hline
        \textbf{Parameter} & 
        \makecell{\textbf{Injected}\\\textbf{value}} & 
        \makecell{\textbf{Recovered}\\\textbf{value}} & 
        \makecell{\textbf{Prior}\\\textbf{(uniform)}} & 
        \makecell{\textbf{Uncertainty}\\\textbf{(FIM) (\%)}} & 
        \makecell{\textbf{Uncertainty}\\\textbf{(NS) (\%)}} \\ \hline
        $N_\mathrm{acc} / 10^{-15}$ & $3.00$   & $3.00$    & $(0,\,20)$    & $0.13$  & $0.13$   \\ \hline
        $\delta x / 10^{-12}$       & $8.00$   & $8.00$    & $(0,\,20)$    & $0.009$ & $0.008$  \\ \hline
        $\log_{10} A_1$                    & $-15.4$  & $-15.39$  & $(-18,\,-5)$  & $0.23$  & $0.22$   \\ \hline
        $\alpha_1$                         & $-5.7$   & $-5.68$   & $(-15,\,-3)$  & $0.67$  & $0.64$   \\ \hline
        $\log_{10} A_2$                    & $-6.32$  & $-6.30$   & $(-10,\,5)$   & $0.45$  & $0.42$   \\ \hline
        $\alpha_2$                         & $-6.2$   & $-6.18$   & $(-10,\,-1)$  & $0.53$  & $0.50$   \\ \hline
        $\log_{10} \Omega_\mathrm{ast}$    & $-11.5$  & $-11.49$  & $(-15,\,-8)$  & $0.20$  & $0.19$   \\ \hline
        $\varepsilon$                      & $2/3$    & $0.66$    & $(0,\,1)$     & $3.24$  & $3.15$   \\ \hline
        $\log_{10} \Omega_0$               & $-11.71$ & $-11.71$  & $(-15,\,-5)$  & $0.39$  & $0.39$   \\ \hline
        $\log_{10} (f_p / \rm Hz)$           & $-2.11$  & $-2.10$   & $(-7,\,-0.1)$ & $1.19$  & $1.23$   \\ \hline
    \end{tabular}
    }
    \caption{
    Fiducial values and uniform prior ranges adopted for data simulation and NS analysis, together with the recovered parameter values and relative uncertainties from both FIM and NS. These values are chosen mainly based on Refs.~\cite{Guan:2025idx, Chen:2023zkb, Boileau:2021kiw}. The SW component yields an absolute SNR of $\mathrm{SNR}_a \simeq 97$ and a relative SNR of $\mathrm{SNR}_r \simeq 61$, discussed at the end of Sec.~\ref{vcn}. The two methods show excellent agreement across all parameters. In several cases, the NS uncertainties appear slightly smaller than those predicted by the FIM; this does not violate the Cramér--Rao bound and is most likely a consequence of statistical fluctuations inherent in the simulated data. A further advantage of NS is that it directly provides the BF; for the SW component, we obtain $\ln \mathrm{BF} \simeq 41$, indicating decisive evidence in favor of the signal model.
    }
    \label{kgb}
\end{table}


\begin{table}[t]
    \centering
    \resizebox{\textwidth}{!}{%
    \begin{tabular}{|c|c|c|c|c|c|c|c|c|c|c|c|}
    \hline
    $\log_{10}\Omega_0$
     & $-12.0$ & $-11.9$ & $-11.8$ & $-11.7$ & $-11.6$
     & $-11.5$ & $-11.4$ & $-11.3$ & $-11.2$ & $-11.1$ & $-11.0$ \\ \hline
    $\mathrm{SNR}_a$
     & $49.6$ & $62.4$ & $78.6$ & $98.9$ & $124.5$
     & $156.8$ & $197.4$ & $248.5$ & $312.8$ & $393.8$ & $495.8$ \\ \hline
    $\mathrm{SNR}_r$
     & $31.2$ & $39.2$ & $49.4$ & $62.2$ & $78.2$
     & $98.5$ & $124.0$ & $156.1$ & $196.6$ & $247.4$ & $311.5$ \\ \hline
    $\ln \mathrm{BF}$
     & $7.6$ & $14.7$ & $28.5$ & $40.0$ & $77.4$
     & $100.5$ & $177.6$ & $297.5$ & $479.9$ & $742.1$ & $1198.9$ \\ \hline
    \end{tabular}
    }
    \caption{
    SNR values (including both the absolute $\mathrm{SNR}_a$ and the relative $\mathrm{SNR}_r$; see Sec.~\ref{vcn} for definitions) and $\ln\mathrm{BF}$ for different injected $\Omega_0$ amplitudes. Both quantities increase with signal strength, showing that stronger SW signals not only enhance the detection significance, but also yield greater Bayesian evidence in favor of the signal model.
    }
    \label{tab:placeholder}
\end{table}


For the statistical inference, we adopt the same likelihood and FIM formulations as in~\cite{Guan:2025idx}. The log-likelihood function is given by
\begin{equation}
\begin{aligned}
    \ln \mathcal{L} ={}& - \sum_{\kappa=1}^{N_0} \sum_{k=1}^{N/2}
    \Bigg\{
    \ln \left[
        \frac{\pi^3 T^3 f_s^6
        (S_A + N_A)(S_E + N_E)N_T}{8}
    \right]
    \\
    &+ \frac{2}{T f_s^2}
    \bigg[
        \frac{|\tilde{d}_A^\kappa(f_k)|^2}{S_A(f_k) + N_A(f_k)}
        + \frac{|\tilde{d}_E^\kappa(f_k)|^2}{S_E(f_k) + N_E(f_k)}
    \\
    &\hspace{7.2em}
        + \frac{|\tilde{d}_T^\kappa(f_k)|^2}{N_T(f_k)}
    \bigg]
    \Bigg\}.
\end{aligned}
\end{equation}
Here, $\tilde{d}_a^\kappa(f_k)$ denotes the discrete Fourier transform of the $a$th TDI channel ($a = A, E, T$) in the $\kappa$th data segment. The sums run over the $N_0$ statistically independent segments and over the positive-frequency bins up to the Nyquist frequency, with $N$ denoting the total number of frequency bins in each segment.

The corresponding FIM takes the form
\begin{equation}
\begin{aligned}
    F_{ij} = N_0 \sum_{k=1}^{N/2}
    \Bigg[
        &\frac{2}{\left(S_A(f_k) + N_A(f_k)\right)^2}
        \frac{\partial \left(S_A(f_k) + N_A(f_k)\right)}{\partial \theta_i}
    \\
        &\hspace{3.4em}
        \frac{\partial \left(S_A(f_k) + N_A(f_k)\right)}{\partial \theta_j}
    \\
        &+ \frac{1}{N_T^2(f_k)}
        \frac{\partial N_T(f_k)}{\partial \theta_i}
        \frac{\partial N_T(f_k)}{\partial \theta_j}
    \Bigg],
\end{aligned}
\end{equation}
where $\theta_i$ denotes the model parameters.


\begin{figure}[t]
    \centering
    \includegraphics[width=0.70\linewidth]{figure/simple_sw_getdist.png}
    \caption{
    Comparison of parameter constraints obtained from NS (red) and FIM (blue). The dark and light shaded regions denote the $68\%$ and $95\%$ confidence intervals, respectively. The analysis includes two instrumental noise parameters ($N_{\mathrm{acc}}, \delta x$), four Galactic foreground parameters ($A_1, \alpha_1, A_2, \alpha_2$), two astrophysical background parameters ($\Omega_{\mathrm{ast}}, \varepsilon$), and two SW signal parameters ($\Omega_0, f_{\mathrm{p}}$). The corner plot illustrates that the NS posteriors are in good agreement with the FIM predictions, with mild deviations attributable to fluctuation in sampling and simulation, and weak non-Gaussian effects. The diagonal panels display the one-dimensional marginalized distributions, with dashed vertical lines marking the $1\sigma$ intervals (blue for FIM, red for NS). Inset: the energy density spectra of the individual components: the Galactic foreground $\Omega_\mathrm{DWD}$ (gray), the astrophysical background $\Omega_{\mathrm{GW,ast}}(f)$ (green), the SW signal $\Omega_{\mathrm{SW}}(f)$ (orange), and the detector sensitivity curve (purple). The simulated total data points are shown in blue.
    }
    \label{kdm}
\end{figure}

\section{Parameter Estimation and Phenomenological Studies} \label{vsm}

In this section, we investigate how GW observations---particularly from space-based detectors such as Taiji and LISA---can inform beyond the Standard Model particle-physics models (see \cite{Caprini:2024hue, Guan:2025idx} for examples). In particular, we examine how the detection or nondetection of a SGWB constrains the parameter space of models predicting a strong FOPT in the early Universe. By linking the dynamics of the EWPT to its GW signals, we show that GW observations provide an independent and complementary probe to collider and dark matter experiments, offering additional constraints on scenarios such as the CxSM.

\FloatBarrier

\subsection{Measurements of GW spectrum parameters}

To assess the detectability of the SW signal sourced by a cosmological FOPT, we employ both Bayesian inference via NS and the FIM approach. These methods offer complementary perspectives: NS reconstructs the full posterior distributions, capturing non-Gaussianity and parameter degeneracies, while the FIM provides a fast, approximate forecast of uncertainties under the Gaussian-likelihood assumption.

We perform NS to derive credible intervals and posteriors for $(\Omega_0, f_{\mathrm{p}})$ under various observational setups. In parallel, the FIM is used to estimate the expected sensitivity and to identify possible degeneracies or nonlinear effects that may not be captured by the Gaussian approximation.

Synthetic datasets are generated assuming the fiducial values listed in Table~\ref{kgb}. Bayesian parameter estimation is then carried out using NS with uniform priors, chosen to cover realistic ranges following \cite{Guan:2025idx, Chen:2023zkb}. The inference includes instrumental noise parameters ($N_{\mathrm{acc}}, \delta x$), foreground and background parameters ($A_1,\allowbreak\ \alpha_1,\allowbreak\ A_2,\allowbreak\ \alpha_2,\allowbreak\ \Omega_{\mathrm{ast}},\allowbreak\ \varepsilon$), and the cosmological signal parameters ($\Omega_0, f_{\mathrm{p}}$). The resulting posteriors are compared with the confidence ellipses obtained from the FIM.

Figure~\ref{kdm} shows the NS results: the red contours denote the $68\%$ and $95\%$ credible regions obtained from NS, while the blue ellipses correspond to the FIM predictions. The two approaches exhibit good overall agreement, with small deviations in higher-dimensional projections arising from fluctuations in simulation and sampling and mild departures from Gaussianity.

The NS analysis, performed using \texttt{bilby.dynesty}, yields the recovered parameter values, associated uncertainties, and the BF summarized in Table~\ref{kgb}. For the SW component, we obtain $\ln\mathrm{BF} \simeq 41$, which indicates a strong statistical preference for all models that include the SW contribution over the hypothesis without it.

\begin{figure}[t]
    \centering
    \includegraphics[width=0.75\linewidth]{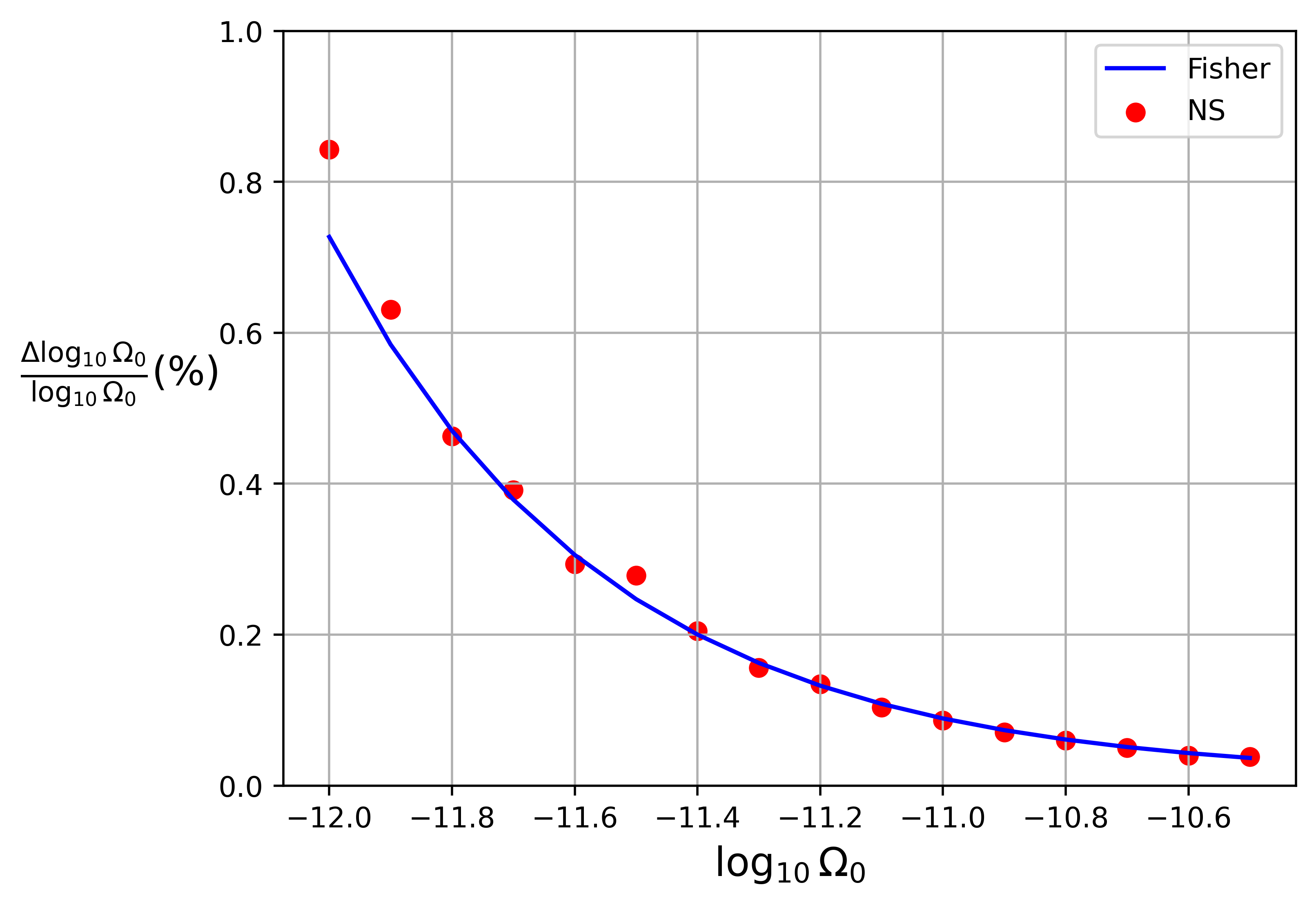}
    \caption{
    Relative uncertainty on $\Omega_0$ obtained from NS (red points) compared with the FIM prediction (blue line). As expected, the uncertainty decreases for larger $\Omega_0$, reflecting the improvement in parameter precision with increasing SW signal strength.
    }
    \label{gbv}
\end{figure}

To further illustrate the dependence of parameter precision on the signal amplitude, we perform NS analysis for a sequence of injected $\Omega_0$ values, compute the corresponding relative uncertainties, and compare them with the FIM predictions. The results are shown in Fig.~\ref{gbv}. We also list the associated SNR and BF values for these injections in Table~\ref{tab:placeholder}, demonstrating explicitly how both quantities scale with increasing signal strength.

\FloatBarrier

\subsection{Measurements of model parameters and Higgs self-couplings}

We perform a comprehensive scan over the CxSM parameter space, characterized by five key input parameters, with the benchmark point given by

\[
    \begin{aligned}
        v_s &= 55.60~\mathrm{GeV}, \,
        m_s = 71.20~\mathrm{GeV}, \,
        m_A = 1832~\mathrm{GeV}, \, \\
        \theta &= 0.45, \,
        a_1 = -2.52\times10^{5}~(\mathrm{GeV})^3.
    \end{aligned}
\]
This benchmark leads to the thermodynamic parameters
\[
    \frac{\beta}{H_n} = 593, \quad 
    T_n = 43.47~\mathrm{GeV}, \quad 
    \alpha = 0.44,
\]
which together describe a strong FOPT capable of producing an observable GW signal. The geometric parameters $(\Omega_0, f_{\mathrm{p}})$ of the SW component listed in Table~\ref{kgb} are derived directly from these thermodynamic values, yielding an absolute SNR of $\mathrm{SNR}_a \simeq 97$ and a relative SNR of $\mathrm{SNR}_r \simeq 61$.

\begin{figure}[t]
    \centering
    \includegraphics[width=0.45\linewidth]{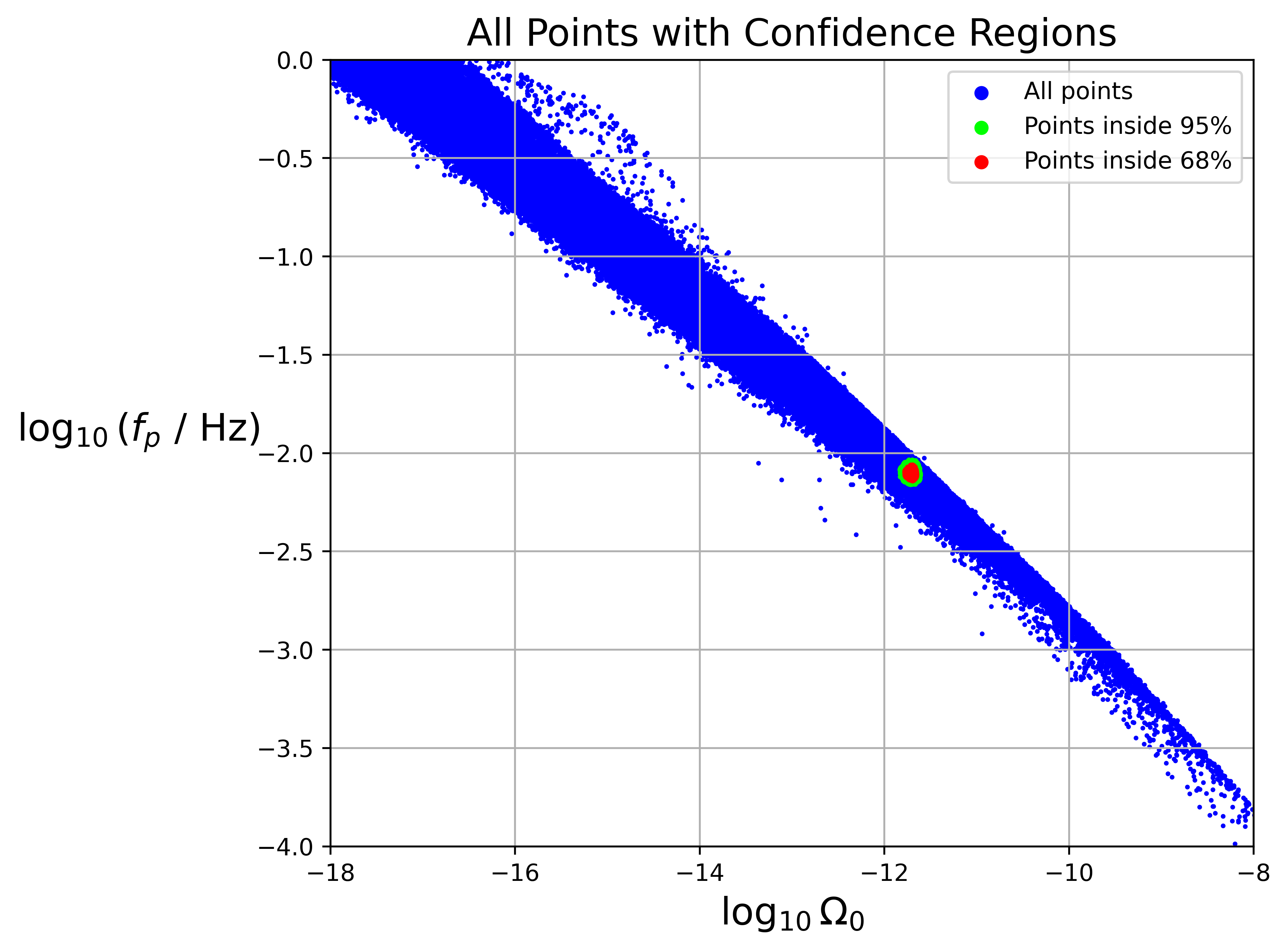}
    \includegraphics[width=0.46\linewidth]{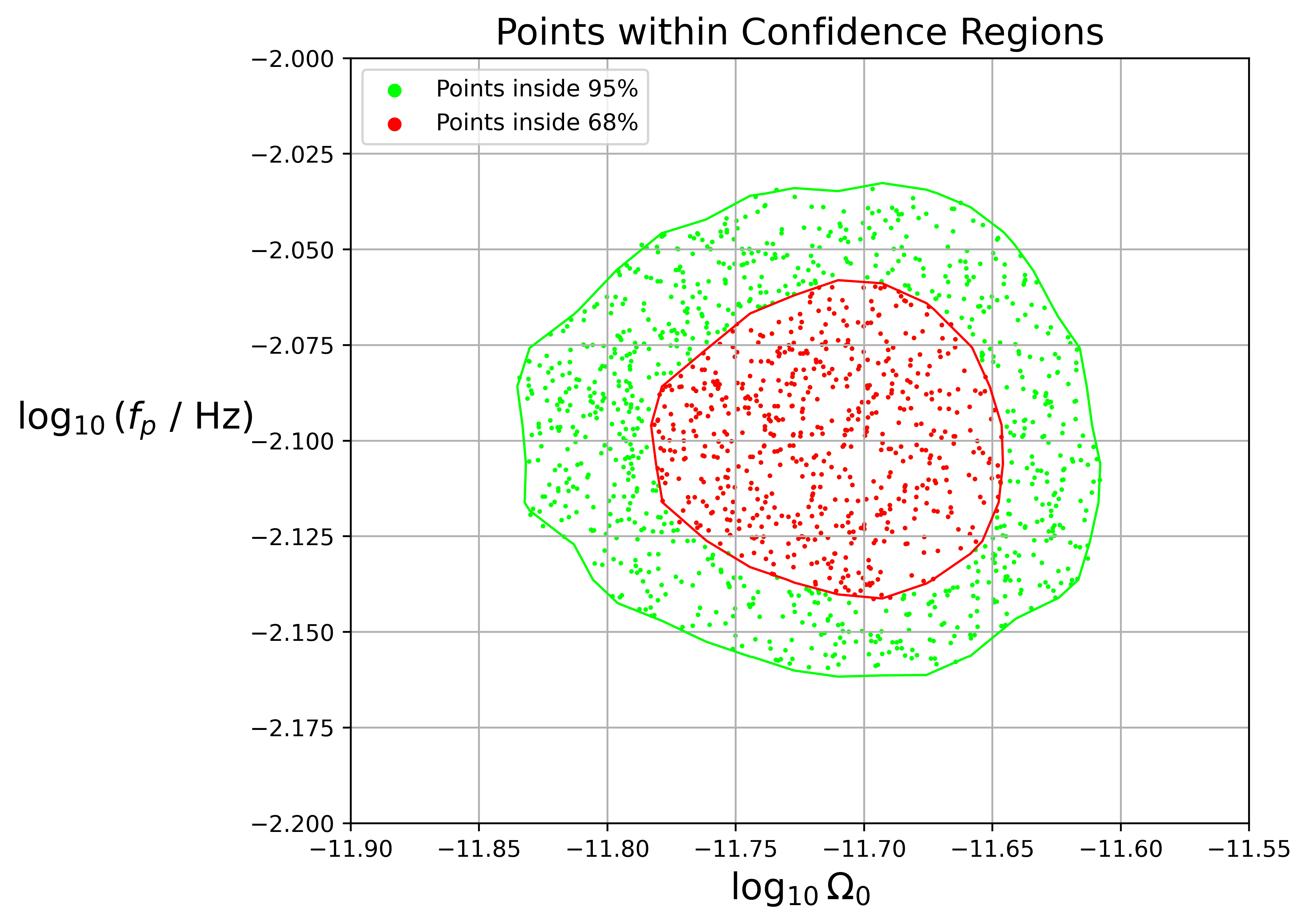}
    \caption{
    Constraints on the geometric parameters $\log_{10}\Omega_0$ and $\log_{10} (f_p / \rm Hz)$ obtained from NS. Blue points represent the full set of parameter combinations from the CxSM scan, while red and green points correspond to those falling within the 68\% and 95\% confidence regions, respectively. Left: the complete distribution of sampled points. Right: provides an enlarged view highlighting the region favored by the data.
    }
    \label{geom_constraints}
\end{figure}

For each sampled point in the CxSM parameter space, we compute the corresponding geometric parameters $(\Omega_0, f_{\mathrm{p}})$ that characterize the SW contribution to the GW spectrum. These results are then compared with the confidence regions obtained from both the FIM and NS analysis, presented in Fig.~\ref{geom_constraints}.

By exploiting the established mapping between the geometric parameters and the CxSM model parameters, we propagate the constraints on $(\Omega_0, f_{\mathrm{p}})$ back to the underlying model space. This procedure allows us to identify the regions of the CxSM parameter space that are most compatible with a potential GW detection. The corresponding NS results are presented in Fig.~\ref{lfv}.

\begin{figure}[t]
    \centering
    \includegraphics[width=0.9\linewidth]{figure/mcmc_constrain_3.png}
    \caption{
    Constraints on the CxSM parameters inferred from the NS posterior distributions. Blue points denote the full set of scanned parameter points, while red and green points correspond to those lying within the 68\% and 95\% credible regions, respectively, as determined from the geometric parameter posteriors. The distributions illustrate how GW observations can significantly narrow the viable regions of the scalar potential parameter space.
    }
    \label{lfv}
\end{figure}

\begin{figure}[t]
    \centering
    \includegraphics[width=0.9\linewidth]{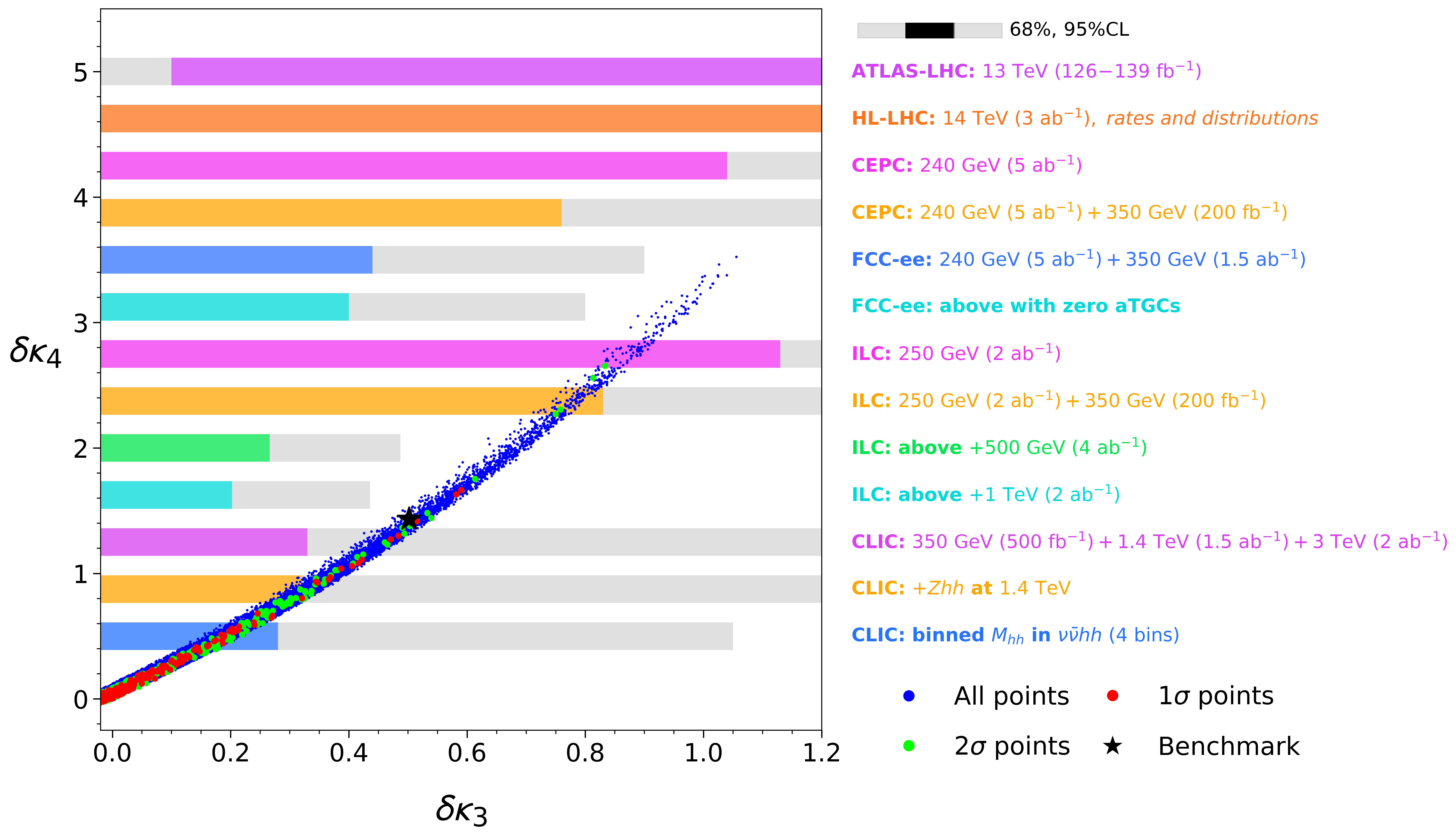}
    \caption{
    Constraints on the Higgs self-coupling deviations $\delta\kappa_3$ and $\delta\kappa_4$ inferred from the Bayesian analysis of a stochastic GW signal in the CxSM. Each point corresponds to a viable model configuration capable of producing a GW spectrum characterized by $(\Omega_0, f_{\mathrm{p}})$. Blue points denote the full CxSM scan, while red and green points indicate the subsets consistent with the $1\sigma$ and $2\sigma$ confidence regions, respectively, in the spectral parameter plane $(\log_{10}\Omega_0,\,\log_{10} (f_p / \rm Hz))$. The black filled star marks the benchmark model used for signal injection. Horizontal bars show the projected $68\%$ and $95\%$ sensitivities to $\delta\kappa_3$ at future colliders such as the HL-LHC, CEPC, ILC, FCC, and CLIC~\cite{DiVita:2017vrr}. This comparison highlights the complementarity between GW inference and collider measurements in probing the scalar potential.
    }
    \label{lmb}
\end{figure}

The posterior distributions obtained from the NS analysis can be mapped onto the Higgs self-coupling deviations, $\delta\kappa_3$ and $\delta\kappa_4$, using the relations discussed in Sec.~\ref{jvk}. Figure~\ref{lmb} shows the resulting correlations between these two couplings in the CxSM. Each point corresponds to a phenomenologically viable CxSM parameter configuration that gives rise to a stochastic GW spectrum characterized by a specific pair of spectral parameters $(\Omega_0, f_{\mathrm{p}})$. Blue points represent the full set of scanned CxSM samples, while the red and green points denote the subsets consistent with the $1\sigma$ and $2\sigma$ confidence regions, respectively, in the spectral parameter plane $(\log_{10}\Omega_0,\,\log_{10} (f_p / \rm Hz))$. The black filled star marks the benchmark model used to generate the mock GW signal.

As illustrated in the figure, the inferred distributions exhibit clear correlations between $\delta\kappa_3$ and $\delta\kappa_4$, reflecting how variations in the scalar potential parameters impact the shape and amplitude of the GW spectrum. Because planned experiments have little sensitivity to the quartic deviation $\delta\kappa_4$~\cite{ATLAS:2024xcs}, the vertical placement of the bands in $\delta\kappa_4$ is purely illustrative and does not reflect any physical constraints. Nevertheless, the noticeable narrowing of the red and green regions compared with the full scan demonstrates that GW-based inference can meaningfully restrict the viable self-interaction structure of the scalar potential, even in parameter directions that remain poorly constrained by collider measurements. 

For comparison, the horizontal bars in Fig.~\ref{lmb} indicate the projected $68\%$ and $95\%$ sensitivities to $\delta\kappa_3$ at future collider facilities, including the High-Luminosity LHC (HL-LHC), Circular Electron-Positron Collider (CEPC), International Linear Collider (ILC), Future Circular Collider (FCC), and Compact Linear Collider (CLIC)~\cite{DiVita:2017vrr,ATLAS:2025nda}. This highlights the strong complementarity between GW observations and collider probes in exploring extended Higgs sectors.

Since the constraints on the Higgs self-coupling parameters obtained above are derived from a single injected benchmark signal, it is important to assess how sensitive these results are to the assumed signal strength. To provide a more general and transparent understanding of this dependence, we therefore consider two additional injected benchmarks, representing a more pessimistic and a more optimistic signal scenario.

By comparing the resulting constraints on the Higgs self-couplings in these three cases, we explicitly illustrate how the precision of GW-based inference scales with the SNR. The corresponding results are presented and discussed in the Appendix.

\FloatBarrier

\section{Conclusion}\label{qmg}

In this work, we have presented a comprehensive study of cosmological phase transitions and their GW signals within the framework of the CxSM. By linking the scalar potential parameters to the thermodynamic quantities of the FOPT and subsequently to the spectral properties of the resulting SGWB, we have established a self-consistent bridge between particle physics and GW observations.

Using a Taiji-like detector response model, we constructed a frequency-domain likelihood that incorporates instrumental noise and astrophysical foregrounds and backgrounds and applied both FIM and NS analysis. The two approaches yield consistent results, demonstrating that Taiji possesses the sensitivity required to detect millihertz GWs from strong FOPT and to precisely recover the key spectral parameters $(\Omega_0, f_{\mathrm{p}})$.

We further propagated these constraints onto the CxSM parameter space and mapped them to the Higgs self-coupling deviations $(\delta\kappa_3, \delta\kappa_4)$. The results show that GW-based inference can place meaningful restrictions on the scalar potential and its self-interaction structure, providing a complementary avenue to collider measurements such as those anticipated at the HL-LHC, CEPC, ILC, FCC, and CLIC.

Overall, this work highlights the strong synergy between future space-based GW detectors and collider experiments in probing electroweak-scale new physics. The methodology developed here offers a general and flexible framework for connecting theoretical models of phase transitions with realistic GW data analysis, paving the way toward a deeper understanding of the Higgs sector through multimessenger cosmology.

\acknowledgments

We would like to thank Ju Chen, Ming-Hui Du, and Chang Liu for helpful discussions. This work is supported by the startup fund provided by the University of Chinese Academy of Sciences and by the National Science Foundation of China (NSFC) under Grants No. 12547104 and No. 12475109. L. B. is supported by the National Natural Science Foundation of China (NSFC) under Grants No. 12547101, No. 12322505, and No. 12075041. L. B. also acknowledges the Chongqing Natural Science Foundation under Grant No. CSTB2024NSCQ-JQX0022 and Chongqing Talents: Exceptional Young Talents Project No. cstc2024ycjh-bgzxm0020. Y. W. is supported by NSFC under Grant No. 12305112.

\section*{Data availability}

The data that support the findings of this study are available from the authors upon reasonable request.

\appendix
\section{Dependence of Higgs Self-Coupling Constraints on Signal Strength}\label{wov}

To further illustrate how the strength of the injected GW signal affects the inferred constraints on the Higgs self-couplings, we consider two additional benchmark scenarios beyond the fiducial case discussed in the main text. These benchmarks are chosen to represent a pessimistic signal with relatively low SNR and an optimistic signal with significantly enhanced detectability.

For the pessimistic benchmark, we choose
\[
    \begin{aligned}
        v_s &= 49.87~\mathrm{GeV}, \,
        m_s = 66.58~\mathrm{GeV}, \,
        m_A = 1591~\mathrm{GeV}, \, \\
        \theta &= 0.44, \,
        a_1 = -1.96\times10^{5}~(\mathrm{GeV})^3.
    \end{aligned}
\]
This benchmark yields
\[
    \log_{10} \Omega_{0} = -12.30, \quad \log_{10} (f_p / \rm Hz) = -1.85,
\]
corresponding to an absolute SNR of $\mathrm{SNR}_a = 10.1$ and a relative SNR of $\mathrm{SNR}_r = 7.3$.

For the optimistic benchmark, we choose
\[
    \begin{aligned}
        v_s &= 44.81~\mathrm{GeV}, \,
        m_s = 66.98~\mathrm{GeV}, \,
        m_A = 394~\mathrm{GeV}, \, \\
        \theta &= 0.40, \,
        a_1 = -1.72\times10^{5}~(\mathrm{GeV})^3.
    \end{aligned}
\]
This benchmark yields
\[
    \log_{10} \Omega_{0} = -11.2, \quad \log_{10} (f_p / \rm Hz) = -2.26,
\]
with $\mathrm{SNR}_a = 425$ and $\mathrm{SNR}_r = 246$.

\begin{figure}[t]
    \centering
    \includegraphics[width=0.9\linewidth]{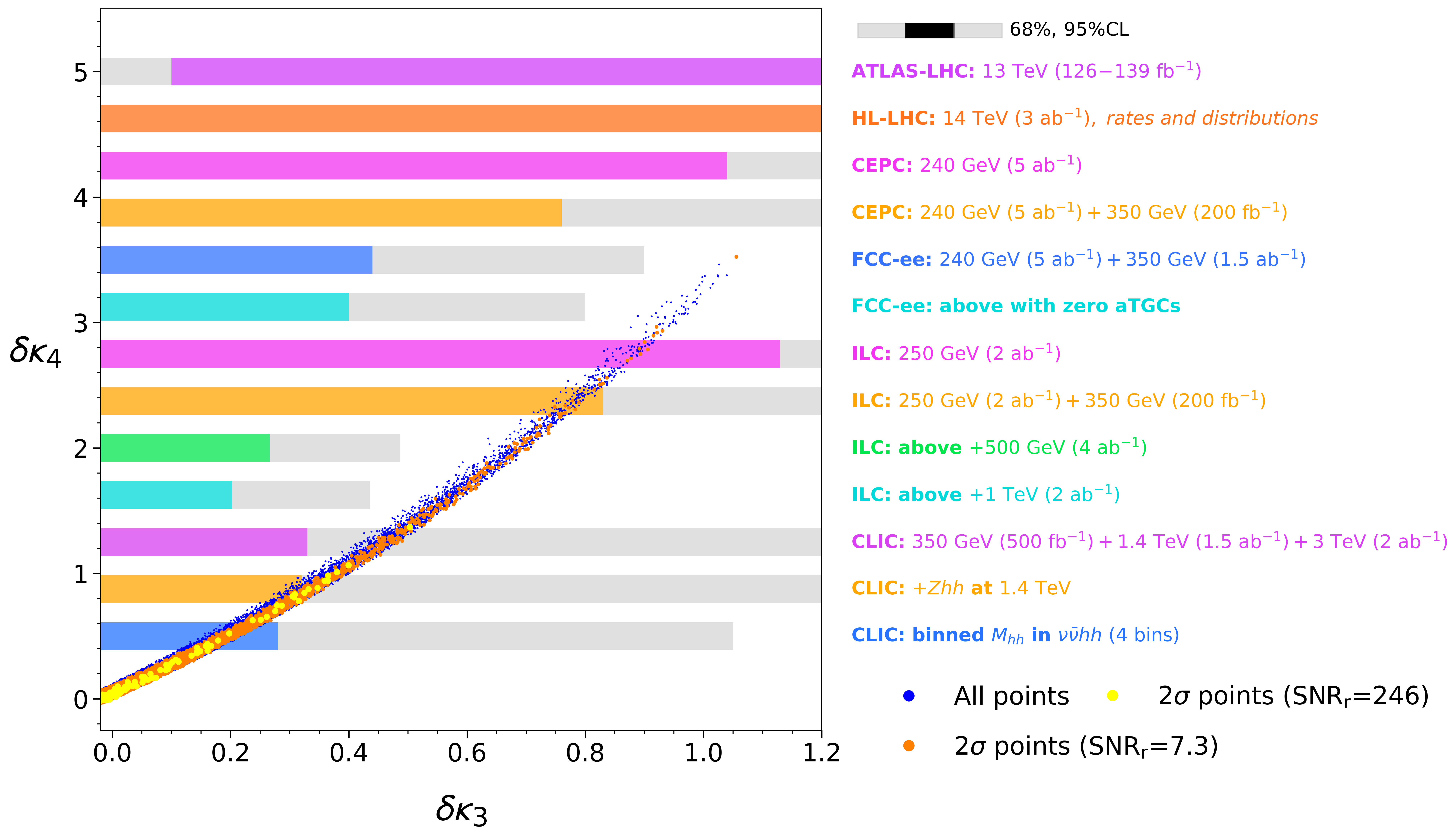}
    \caption{
    Constraints on the Higgs self-coupling deviations $\delta\kappa_3$ and $\delta\kappa_4$ for different injected GW signal strengths. Blue points denote the full CxSM parameter scan. Orange points correspond to the pessimistic benchmark with $\mathrm{SNR}_r = 7.3$, while purple points correspond to the optimistic benchmark with $\mathrm{SNR}_r = 246$, both obtained from the $2\sigma$ FIM confidence regions. The comparison illustrates the strong dependence of the inferred Higgs self-coupling constraints on the SNR of the GW signal.
    }
    \label{wgv}
\end{figure}

The resulting constraints on the Higgs self-coupling deviations are shown in Fig.~\ref{wgv}. The blue points represent the full set of scanned CxSM parameter points, identical to those shown in Fig.~\ref{lmb}. The orange and purple points correspond to parameter points lying within the $2\sigma$ FIM confidence regions derived from the pessimistic and optimistic benchmarks, respectively.

In this appendix, we employ FIM forecasts rather than NS. As demonstrated in Fig.~\ref{kdm}, the FIM predictions provide an accurate approximation to the NS results in the regimes considered here. As expected, benchmarks with higher SNR lead to substantially tighter constraints on the Higgs self-coupling parameters, while weaker signals yield correspondingly looser bounds.

\FloatBarrier

\bibliographystyle{JHEP}
\bibliography{refer}

\end{document}